\documentclass[a4paper,11pt]{article}
\usepackage{pos}
\usepackage{graphicx}
\usepackage{subcaption}
\usepackage{amsmath}
\newcommand{\dif}{\textrm{d}}
\def\th{\theta}

\title{Jet fragmentation function in heavy-ion collisions}

\author*[a]{P.~Caucal}
\author[a]{E.~Iancu}
\author[b]{A.H.~Mueller}
\author[a]{G.~Soyez}

\affiliation[a]{Institut de Physique Th\'{e}orique, Universit\'{e} Paris-Saclay, CNRS, CEA, F-91191, Gif-sur-Yvette, France}

\affiliation[b]{Department of Physics, Columbia University, New York, NY 10027, USA}

\emailAdd{paul.caucal@ipht.fr}

\abstract{We study the fragmentation function of jets propagating through a
dense quark-gluon plasma within perturbative QCD.
Our results for its nuclear modification factor are in qualitative agreement with the experimental data in Pb+Pb collisions at the LHC. In particular, we reproduce the enhancements seen in the data at both relatively soft and relatively large transverse momenta, with clear physical interpretations. The perturbative predictions however are quite sensitive to the value of the infrared cutoff mimicking the confinement scale, due to the fact that the fragmentation function is not an infrared safe quantity. To remedy this, we propose a new observable --- the (primary) subjet fragmentation function --- which is infrared safe and has features similar to the fragmentation function. We provide predictions for this observable in the vacuum and in heavy-ion collisions that could be tested against the experimental data.}

\FullConference{%
  HardProbes2020\\
  1-6 June 2020\\
  Austin, Texas}

\begin{document}
\maketitle

%

One of the most striking evidence for the formation of a quark-gluon plasma in heavy-ion collisions at RHIC and at the LHC is the suppression of high $p_T$ jets, phenomenon known as jet quenching. If the measurement of such suppression is of fundamental importance \textit{per se}, it is not enough to constrain the various theoretical models for plasma-jet interactions. It is then crucial to confront models with other jet observables. Among them, jet substructure observables are promising probes of the modifications of the partonic cascades produced in the medium \cite{Andrews:2018jcm}.
In this paper, we discuss one such observable: the jet fragmentation function.

The theoretical framework for this study is perturbative QCD (pQCD) and its emerging picture for jet evolution detailed in \cite{Caucal:2018dla,Caucal:2019uvr}. This picture includes two kinds of radiation: standard vacuum-like emissions (VLEs) triggered by the parton virtuality and medium-induced emissions (MIEs) triggered by multiple collisions inside the medium, characterised by its quenching parameter $\hat{q}$. 
%
%
%
We have shown in \cite{Caucal:2018dla} that the partonic cascades can be factorised in three steps: {\tt (i)} a cascade of VLEs {\em inside the
    medium} satisfying $\omega^3\theta^4>2\hat{q}$ and $\theta>\theta_c \equiv
  2/\sqrt{{\hat q}L^3}$ with $\omega$ and $\theta$ the energy and angle of emission and $L$ the jet path length, 
{\tt (ii)} each parton escaping the previous vacuum-like cascade propagates
  through the medium over a distance $L$ and thus sources
  MIEs,
{\tt (iii)} finally all the outgoing partons trigger another cascade of VLEs outside
  the medium in the phase space $t_f=2/\omega\theta^2>L$. The first emission outside the medium is not constrained by angular ordering because of the colour decoherence due to the interactions with the plasma \cite{MehtarTani:2010ma,MehtarTani:2011tz,CasalderreySolana:2011rz}.
The implementation of this picture within a Monte-Carlo (MC) parton shower is straightforward (see \cite{Caucal:2018ofz,Caucal:2019uvr} for detailed descriptions). All the numerical results presented here are obtained using this~MC.

This paper is organised as follows: we first present our results for the ``standard'' fragmentation function defined from the hadrons inside jets. This observable being infrared and collinear (IRC) unsafe, we define a new fragmentation function under better theoretical control in pQCD. This IRC safe fragmentation function is then studied in the vacuum and in nucleus-nucleus collisions.

\section{Nuclear modifications of the standard fragmentation function}

The standard jet fragmentation function $\mathcal{D}(x)$ and its nuclear modification factor $\mathcal{R}(x)$ are defined as
\begin{equation}\label{frag-def}
 \mathcal{D}(x)=\frac{1}{N_{\textrm{jets}}}\frac{\dif N}{\dif x},\qquad \mathcal{R}(x)=\frac{\mathcal{D}^{\textrm{med}}(x)}{\mathcal{D}^{\textrm{vac}}(x)}
 \end{equation}
where $N_{\textrm{jets}}$ is the number of selected jets and $\dif N$ is the number of jet constituents (hadrons) with a momentum fraction $x$ between $x$ and $x+\dif x$. The integral of $\mathcal{D}(x)$ over $x$ is equal to the mean total intrajet multiplicity of hadrons, a quantity which is of course not IRC safe. Thus, the fragmentation function is strongly sensitive to hadronisation corrections
%
or on the unphysical cut-off $k_{\perp,\rm min}$ required to regulate the collinear divergence in a parton level calculation.
\begin{figure}[t] 
  \centering
  \begin{subfigure}[t]{0.48\textwidth}
    \includegraphics[page=1,width=\textwidth]{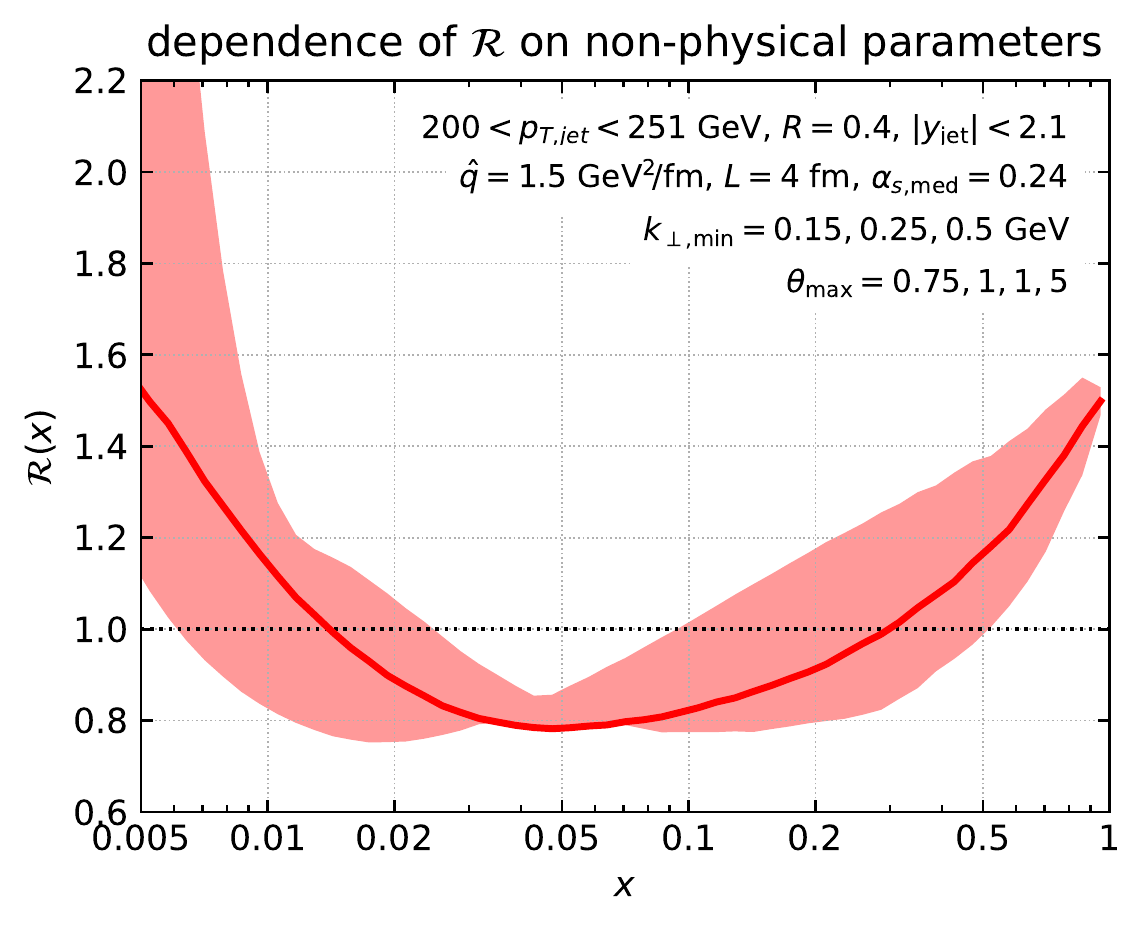}
    \caption{\small Variations in  $\theta_{\rm max}$ and $k_{\perp,\text{min}}$.}\label{Fig:MCunphys} 
  \end{subfigure}
  \hfill
  \begin{subfigure}[t]{0.48\textwidth}
    \includegraphics[page=1,width=\textwidth]{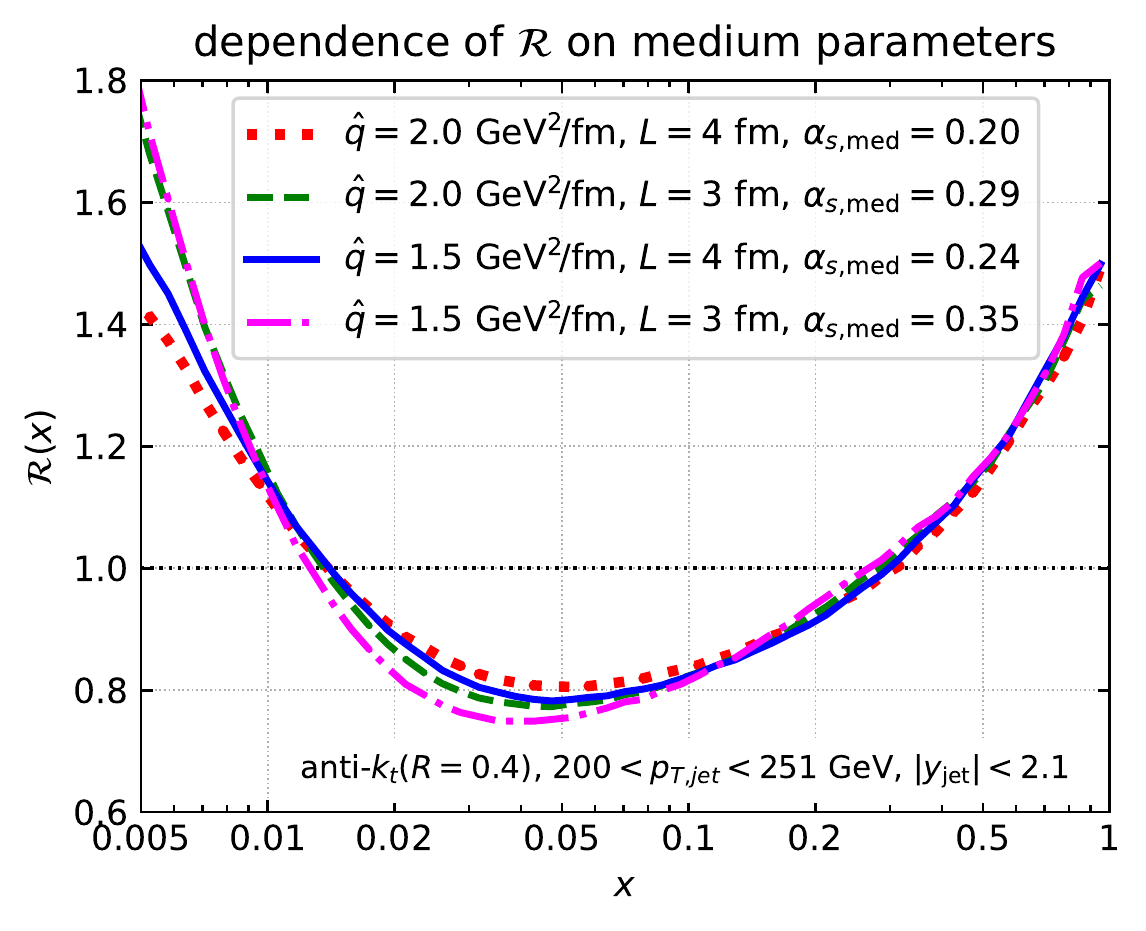}
    \caption{\small Variations in  $\hat{q}$, $L$ and $\alpha_{s,\rm med}$.}
    \label{Fig:MC-pheno}
  \end{subfigure}
  \caption{\small (Left) Variability of our MC results for the ratio $\mathcal{R}(x)$ w.r.t. changes
  in $k_{\perp,\rm min}$ and the maximal branching angle $\th_{\rm max}$. The envelop is dominated by $k_{\perp,\rm min}$ variations. (Right) MC results for the 4 sets of values of
  the medium parameters $\hat{q}$, $L$ and the strong coupling for medium-induced vertices $\alpha_{s,\rm med}$ that provide similarly good descriptions
  of the LHC data~\cite{Aaboud:2018twu} for the nuclear  modification factor for jets $R_{AA}$~\cite{Caucal:2019uvr}.}
  \label{fig-R-param-dependence}
\end{figure}

To illustrate this, we show on Fig.~\ref{fig-R-param-dependence}-left the variability of $\mathcal{R}(x)$ with respect to the cut-off $k_{\perp,\rm min}$ as given by our MC, which does not currently include any hadronisation model. As expected, the envelop of the variations is large. Nevertheless, the general trend of the curve is in qualitative agreement with the recent ATLAS data \cite{Aaboud:2018hpb}. Namely, one observes a significant enhancement of the ratio at small $x$ and for $x$ close to $1$, while for $0.02\lesssim x \lesssim 0.2$, the ratio becomes smaller than~$1$. We point out that since both $x\mathcal{D}^{\textrm{med}}(x)$ and $x\mathcal{D}^{\textrm{vac}}(x)$ are normalized to~1, one cannot have $\mathcal{R}(x)>1$ for all $x$. That said, we now briefly explain the physical mechanisms leading to these enhancements at both ends of the spectrum.

In the small $x$ part, these mechanisms differ significantly from those encountered in the literature~\cite{Casalderrey-Solana:2016jvj,Tachibana:2017syd,KunnawalkamElayavalli:2017hxo}, 
where the increase of soft particle production at
relatively large angles from the jet axis is related to the medium response to the jet propagation. In our picture, soft particle production is dominated by gluons produced outside the medium with $t_f>L$ at the end of the evolution. This evolution is amplified in the presence of a medium w.r.t.\ the vacuum because of \cite{Caucal:2020xad}: (i) the violation of angular ordering by the first emission outside the medium, which
re-opens the angular phase-space for subsequent evolution \cite{Mehtar-Tani:2014yea,Caucal:2018dla}, and (ii) the presence
of relativitely hard MIEs with $\omega\gg\omega_{\rm br}\equiv \alpha_{s}^2\hat{q}L^2$ which remain inside the jet and can further radiate VLEs outside the medium~\cite{Caucal:2019uvr}. (The multiple branching scale $\omega_{\rm br}$ is the typical scale for energy loss at large angles, i.e.\ MIEs deviated out of the jet cone \cite{Blaizot:2015lma}.)

For $x$ close to 1, we have shown in \cite{Caucal:2020xad} that the enhancement seen in the ratio is essentially \textit{not} a change in the in-medium fragmentation process but rather a bias toward hard fragmenting jets induced by the steeply falling jet spectrum.
This spectrum favors jets that lose less energy. In hard-fragmenting jets, the leading parton\footnote{Most likely a quark, since quark jets have a harder core and lose less energy than gluon jets \cite{Spousta:2015fca,Caucal:2020xad}.} sourcing the jet must not have radiated too many partons in order to have a final $x$ fraction close to 1. As the intrajet multiplicity is positively correlated with the energy loss \cite{Caucal:2019uvr}, hard fragmenting jets lose less energy than average jets, hence are favored in nucleus-nucleus collisions.
This argument is actually very general as it applies to many theoretical models \cite{KunnawalkamElayavalli:2017hxo,Rajagopal:2016uip,Casalderrey-Solana:2016jvj,Casalderrey-Solana:2018wrw}. It also explains the strong correlation between the nuclear modification factor for jets $R_{AA}$ and the $x\sim 1$ region of the fragmentation function, since both quantities are controlled by energy loss effects. This correlation is apparent on Fig.~\ref{fig-R-param-dependence}-right: the 4 sets of physical parameters in our calculations that all give a good description of $R_{AA}$ \cite{Caucal:2019uvr} scale identically for $x\sim1$ whereas the degeneracy is lifted for $x\ll 1$. We conclude that by studying the soft sector in the fragmentation of jets, one can better constrain theoretical models that have been tuned to yield a good description of $R_{AA}$.


\section{Fragmentation into subjets}

In this section, we first define an IRC safe fragmentation function relying on subjets and we calculate it at leading logarithmic (LL) accuracy in the vacuum. Then, we argue that this new observable is sensitive to the same medium effects while being resilient to non-perturbative physics.

\paragraph{Definition and LL result in the vacuum.}The basic idea of this observable is to replace the unphysical cut-off $k_{\perp,\rm min}$ by a parameter of the observable, $k_{\perp,\rm cut}$ which is chosen much larger than the confinement scale $\Lambda_{\rm QCD}$. For a given jet with transverse momentum $p_{T,\textrm{jet}}$, we decluster the jet using the Cambridge/Aachen algorithm \cite{Dokshitzer:1997in,Wobisch:1998wt} following the hardest branch. This gives a list of subjets. In this list, we keep only those which have a transverse momentum with respect to the hard branch larger than $k_{\perp,\textrm{cut}}$. The subjet fragmentation  is then defined as the number $\dif N_{\textrm{sub}}$ of subjets with \textit{splitting fraction} $z<1/2$ between $z$ and $z+\dif z$ normalised by the total number of selected jets~\cite{Caucal:2020xad}:
\begin{equation}
 \mathcal{D}_{\textrm{sub}}(z)\equiv\frac{1}{N_\textrm{jets}}\frac{\dif N_{\textrm{sub}}}{\dif z}
\end{equation}
This definition is close to the Iterated Soft Drop multiplicity differential in $z$ \cite{Frye:2017yrw}. 

To calculate this quantity in the vacuum, it is convenient to introduce the generating functional of the exclusive probability distribution $P _i^{(n)}(z_1,...,z_n)$ of $n$ subjets with splitting fractions $z_1$,...,$z_n$ in a jet with transverse momentum $p_T$, opening angle $\theta$ and ``flavor'' $i\in\{q,g\}$:
 \begin{equation}\label{Zjet}
 Z_i(p_T,\theta|u(z))=\sum_{n=1}^{\infty}\int \dif z_1...\dif z_n\, u(z_1)...u(z_n)P_i^{(n)}(z_1,...,z_n)
\end{equation}
The fragmentation function into subjets $\mathcal{D}_{i,\rm sub}(z|p_T,R)$ of a $i$-jet with opening angle $R$ and transverse momentum $p_T$ is then obtained by a functional differentiation of $Z_i(p_T,R|u(z))$ with respect to the probing function $u(z)$, at $u=1$.
From the Markovian property of the branching process ordered in angles,  $Z_i$ satisfies the following differential equation in the vacuum:
\begin{equation}\label{Z-primary}
\frac{\partial Z^{\rm vac}_{i}(p_T,\theta|u)}{\partial \log(\theta)}=\sum_{(a,b)}\int_0^{1/2}\dif z\,\frac{\alpha_s(k_\perp)}{\pi}\Theta_{\rm cut}\Phi_i^{ab}(z)\left(u(z)Z^{\rm vac}_b((1-z)p_T,\th|u)-Z^{\rm vac}_i(p_T,\th|u)\right)
\end{equation}
where $\Phi_i^{ab}(z)$ are the unregularised DGLAP splitting functions (the sum runs over all distincts pairs of partons $(a,b)$) and $\Theta_{\rm cut}$ is a step function enforcing $k_\perp>k_{\perp,\rm cut}$. Eq.~\ref{Z-primary} includes corrections associated with the recoil of the hard branch. At LL accuracy, one can ignore such contributions by setting $Z_b((1-z)p_T,\th|u)\simeq Z_b(p_T,\th|u)$ and one can approximate $\Phi_i^{ab}(z)\simeq 2C_i/z$ and $k_\perp\simeq zp_T\th$ in the argument of $\alpha_s$. At this stage, it is possible to solve exactly the corresponding evolution equation for $\mathcal{D}_{\rm sub}(z|p_T,\theta)$ with initial condition $\mathcal{D}_{\rm sub}(z|p_T,0)=0$. Furthermore, if one neglects quark-gluon mixing\footnote{Even though it formally matters at LL accuracy.}, one finds a good approximation of the LL result under the simple form:
\begin{equation}\label{frag-sub-vac}
\mathcal{D}^{\rm vac}_{i,\rm sub}(z|p_T,R)\simeq \frac{2C_i}
{\pi}\int_{0}^R\frac{\dif\theta}{\theta}\,
\frac{\alpha_s(zp_T\th)}{z}\Theta(zp_T\th-k_{\perp,\rm cut})
\end{equation}
This formula shows explicitly how the parameter $k_{\perp,\rm cut}\gg \Lambda_{\rm QCD}$ plays also the role of an infrared regulator in the $\th$ integral. It makes manifest the connection between $\mathcal{D}_{\rm sub}(z)$ and the primary Lund plane density integrated over all angles with $z\simeq k_\perp/(\theta p_{T,\textrm{jet}})$ fixed \cite{Dreyer:2018nbf,Lifson:2020gua}.  Finally, the resummed result \eqref{frag-sub-vac} should be matched with a fixed order calculation for quantitative comparisons with data in $pp$ collisions.

\begin{figure}[t] 
  \centering
  \includegraphics[page=2,width=0.48\textwidth]{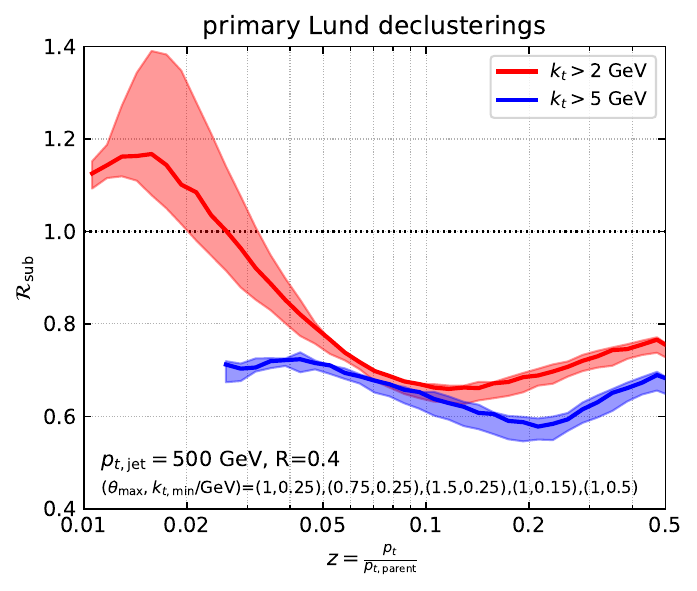}
  \hfill
   \includegraphics[page=5,width=0.48\textwidth]{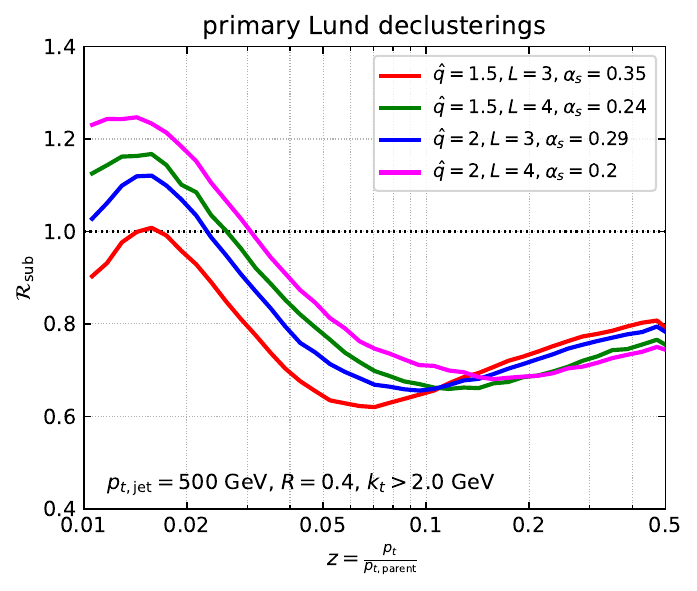}
  \caption{\small Variability of our MC results for the ratio $\mathcal{R}_{\rm sub}(z)$ w.r.t.\ changes
  in $k_{\perp,\rm min}$ and  $\th_{\rm max}$ (left) and w.r.t\ the medium parameters (right) that reproduce the ATLAS $R_{AA}$ ratio as in Fig.~\ref{Fig:MC-pheno}. For the latter figure, $k_{\perp,\textrm{cut}}=2$~GeV and the unphysical parameters are fixed to 
$\theta_{\textrm{max}}=1$ and $k_{\perp,\rm min}=250$~MeV.
   }
\label{Fig:subjet-med} 
\end{figure}

\paragraph{Monte-Carlo results in Pb+Pb collisions.}We now discuss the nuclear modification factor $\mathcal{R}_{\rm sub}(z)$ defined as in \eqref{frag-def} and calculated with our MC. First of all, being IRC safe, $\mathcal{R}_{\rm sub}(z)$ is much less sensitive to $k_{\perp,\rm min}$ as shown Fig.~\ref{Fig:subjet-med}-left. The larger $k_{\perp,\rm cut}$ is, the more this sensitivity is reduced. On Fig.~\ref{Fig:subjet-med}-right, the ratio $\mathcal{R}_{\rm sub}(z)$ is shown for the 4 sets of parameters that fit $R_{AA}$ measured by ATLAS~\cite{Aaboud:2018twu}. The reduction at large $z\sim0.5$ is again essentially a consequence of the normalisation factor $N_{\rm jets}$ \cite{Caucal:2020xad}. Jets with $z\sim 0.5$ and $k_\perp>k_{\perp,\rm cut}$ are typically two-prongs jets with angular separation larger than the coherence angle $\th_c\propto(\hat{q}L^3)^{-1/2}$, and thus lose \textit{more} energy than average jets \cite{Caucal:2019uvr,Casalderrey-Solana:2019ubu}. They are disfavored by the initial jet spectrum and suppressed in Pb-Pb collisions. On the other hand, the increasing behaviour of the ratio as $z$ decreases is mainly a consequence of intrajet semi-hard MIEs \cite{Caucal:2020xad}. Such emissions source subjets that are finally captured in the declustering procedure performed to calculate $\mathcal{D}_{\textrm{sub}}^{\rm med}(z)$. As $k_{\perp,\rm cut}$ is much larger than the scale $2/(LR)$, $\mathcal{D}_{\textrm{sub}}^{\rm med}(z)$ probes only a tiny portion of the ``outside'' region of phase space with $t_f>L$. Angular ordering violation has therefore a less pronounced effect than for the standard ratio $\mathcal{R}(x)$.

To conclude, we point out that $k_{\perp,\rm cut}$ has a determinant role in heavy-ion collisions. It must not be too small to avoid large hadronisation corrections. However, if it is too large, the interesting medium effects in the relatively soft sector highlighted above are not captured. In that respect, substructure observables relying on dynamical grooming is another interesting possibility \cite{Mehtar-Tani:2019rrk,Soto-Ontoso:2020ola}.


\paragraph{Acknowledgements} The work of P.C.,  E.I. and G.S. is supported in part by the Agence Nationale de la Recherche project 
 ANR-16-CE31-0019-01.   The work of A.H.M.
is supported in part by the U.S. Department of Energy Grant \# DE-FG02-92ER40699. 
\bibliographystyle{JHEP}
\bibliography{biblio}
\end{document}